\documentclass[12pt]{article}
\usepackage{amsfonts,amsmath,amssymb,amsthm,bbold}
\usepackage{scalerel}[2016/12/29]
\usepackage{epsfig,graphicx}
\usepackage{tikz}
\usepackage{float}
\usepackage{braket}
\usepackage[nosort]{cite}
\usepackage{url,hyperref}
\usetikzlibrary{decorations,arrows}
\usetikzlibrary{decorations.markings}
\usetikzlibrary{shapes.geometric}

\newcommand{\bb}{\hskip -0.1cm}

\newcommand{\bm}{\bb - \bb}

\def\bb{\hskip -0.5mm}
\def\be{\begin{equation}}
\def\ee{\end{equation}}
\def\bea{\begin{eqnarray}}
\def\eea{\end{eqnarray}}

\def\half{{\textstyle {1\over 2}}}

\def\rme{\mathrm{e}}

\def\rme{\mathrm{e}}

\begin{document}

\title{\bf{Microscopic inclusion statistics in a discrete 1-body spectrum}}

\date{\today}

\author{St\'ephane Ouvry$^{*}$  \and Alexios P. Polychronakos$^{\dagger}$}

\maketitle

\begin{abstract}
We present the microscopic formulation of inclusion statistics, a counterpoint to exclusion
statistics in which particles tend to coalesce more than ordinary bosons. 
We derive the microscopic occupation multiplicities of 1-body quantum states and show that they factorize into
a product of clusters of neighboring occupied states with enhanced statistical weights. Applying this statistics
to a one-dimensional  gas of particles in a harmonic well leads to a Calogero-like $n$-body inclusion spectrum with
interesting physical properties.
\end{abstract}

\noindent
* LPTMS, CNRS,   Universit\'e Paris-Saclay, B\^at Pascal, 91400 Orsay Cedex, France\\
${ }$\hskip 0.38cm{\it stephane.ouvry@universite-paris-saclay.fr}

\noindent
$\dagger$ Physics Department, the City College of New York, New York, NY 10031, and\\
${ }$\hskip 0.35cm The Graduate Center of CUNY, New York, NY 10016, USA
\\ \noindent
${ }$\hskip 0.38cm{\it apolychronakos@ccny.cuny.edu}


\section{Introduction}

Quantum statistics deviating from Bose-Einstein and Fermi-Dirac statistics have been part of physics for over 70 years. Parastatistics, originally devised to explain the properties of quarks, followed by anyon statistics \cite{LDW,LM,thewir}, as a theoretical possibility in two dimensions, nonabelian statistics in systems with degenerate ground states, and
exclusion statistics \cite{Haldane,ASW,NW} as effective descriptions of some interacting systems, are examples of such generalizations.

In all the above variants of statistics, bosons remain the most ``gregarious" type, and the only ones undergoing
Bose-Einstein condensation at sufficiently low temperatures. In a recent development \cite{usus} we generalized 
the notion of statistics to {\it inclusion} statistics, which extends the statistical spectrum
beyond bosons and enhances the particles' property to coalesce. The qualitatively new feature of inclusion
particles is that they can condense in one space dimension lower than ordinary bosons, opening the
possibility for condensation in planar systems. As such, their properties and possible experimental realization
are matters of physical interest.

Most types of statistics have a microscopic formulation in terms of individual particles.
In a previous work \cite{oulabis} we gave a microscopic formulation of exclusion statistics for an integer statistics
parameter in a nondegenerate discrete 1-body spectrum 
and studied its combinatorial properties. In the present work we extend this formulation to inclusion
statistics and study its relation to exclusion statistics, via a duality relation generalizing  the well-known Bose-Fermi correspondence. The concept of
statistical clusters of particles occupying nearby states {{with non trivial multiplicities}} providing the basic building blocks for
inclusion statistics emerges in this analysis. Small variations in the definition of inclusion statistics,
irrelevant in the thermodynamic limit where the number of relevant 1-body quantum states diverges,
become relevant at the microscopic level.

The paper is organized as follows: in sections 2 and 3 we review the concepts of exclusion and inclusion
statistics in the thermodynamic limit and their duality relation. In section 4 we review the microscopic
formulation of exclusion statistics, and in section 5 we define the  microscopic formulation of inclusion
statistics on a lineal 1-body spectrum (with states ordered on a line) and derive its combinatorial properties.
In section 6 we present a possible
realization of inclusion statistics in a harmonic spectrum and relate it to a Calogero-like model. In section 7 we
conclude with some speculations and connections with other combinatorial topics. Technical issues such as the
formulation of microscopic inclusion statistics on a periodic 1-body spectrum (i.e., with 1-body quantum states on a circle), the derivation
of some mathematical results, and the recovery of the free space limit from a weak potential trap, are presented in
the Appendix.

\section{$g$-exclusion \label{333}}
It has been known since the nineties \cite{Poly0,Das,Wu,Poly}, from the study of quantum systems such as the
lowest Landau level
anyon model or the Calogero model, that thermodynamical equations can be obtained for particles with
generalized statistics described by a statistical parameter $g$ interpolating between the standard $g=0$ 
Bose-Einstein and $g=1$ Fermi-Dirac statistics.
The key information for these $g$-statistics is contained in the $n$-body
cluster coefficients $c_{k,n}(g)$; specifically 
\be \ln {\cal Z}_g=\sum_{n=1}^{\infty} c_{k,n}(g)z^n
\label{00000}
\nonumber\ee
where ${\cal Z}_g$ is the grand partition function for particles with statistical parameter $g$ in $k$ degenerate
1-body quantum states (here taken for convenience at  zero energy) and $z=e^{\beta \mu}$ is the fugacity. In the thermodynamic limit $k\to \infty$ the grand potential $\ln {\cal Z}_g$ scales as $k$,
\be
\ln {\cal Z}_g=\sum_{n=1}^{\infty} c_{k,n}(g)z^n\simeq k \sum_{n=1}^{\infty}c_{n}(g)z^n
\label{00}
\ee
This defines effective single-level cluster coefficients $c_n (g)$ and an effective single-level grand partition function $y$ for $g$-statistics,
\be \sum_{n=1}^{\infty}c_{n}(g)z^n=\ln y\label{s1}\ee
where the single-level cluster coefficients are 
\be c_{n}(g)={1\over n}\prod_{i=1}^{n-1} {i-n g\over i}\label{0}
\ee
This leads to  the single-level grand partition function  $y$  defined by the equation
\be
y^g-y^{g-1}=z
\label{1}\ee

A natural step is to forget about the specific models that led to these statistics and adopt (\ref{00}-\ref{1}) as the definition of $g$-statistics in a degenerate spectrum  for a general statistical parameter $g\ge 0$. For $g\ge 1$ one speaks of exclusion statistics,
which can also
be obtained from an {\it ad hoc} Hilbert space counting argument {\it \`a la} Haldane \cite{Haldane}, where each
particle excludes $g$ states from being occupied by other particles. This counting is in general ill-defined for
fractional $g$, as it can lead to a fractional or even negative number of states, and is thus only an effective
description of the system {in the thermodynamic limit}. However, it becomes well-defined for integer values of $g$.

From now on we will focus on  integer values of $g$, which allows for a well-defined microscopic formulation of
exclusion statistics. In this case, (\ref{1}) has $g$ generally complex solutions
$$
y_1=1+z+\ldots,~~ y_i= \rme^{{\rm i}{(2i-1)\pi \over g-1}} z^{1\over g-1} + \ldots , ~i=2,\ldots,g
$$
of which $y_1 $ is the physical one as defined in (\ref{s1},\ref{0}),  meaning that it is real, equal to 1 when $z=0$, and increasing with $z> 0$.
For example, for $g=2$
\[\small{
y_1={1\over 2} \Big(1 + \sqrt{1 + 4 z}\hskip 0.05cm\Big) = 1+z+\dots~,~~~ 
y_2={1\over 2} \Big(1 - \sqrt{1 + 4 z}\hskip 0.05cm\Big) = -z +\dots}
\]
while for $g=3$ 
\bea  y_i&=&{1\over 3 }- {2\over 3}  \cos\Big({\arcsin(1 + 27  {z/ 2})\over 3} +{ \pi\over 6} +  i {2\pi\over 3}\Big)\label{the3}\eea
for $i=1, 2, 3$ (note that $y_2 \bb=\bb y_3^*$ for $z\bb>\bb0$).
The thermodynamic degenerate grand partition function, which from (\ref{00},\ref{s1}) rewrites  as
\be {\cal Z}_g=y_1^k\;,\label{toto}\ee
can be modified to a form that leads to a valid microscopic interpretation, as explained in detail in the Appendix,
namely
\be
{\cal Z}_g=\sum_{i=1}^g y_i^{2(g-1)+k} \prod_{j \neq i , j=1}^g {1\over y_i - y_j}=\sum_{n=0}^{[(k+g-1)/g]} {(k+(1-g)(n-1))!\over n!(k-(n-1) g-1)!}z^n
\label{linealbis}
\ee
In (\ref{linealbis})
the summation over $n$ is truncated at $[(k+g-1)/g]$, as is natural for $g$-exclusion in a spectrum on the line in which
$[(k+g-1)/g]$ particles exclude all states for additional particles.
Moreover, the degeneracy factor  at order $z^n$, i.e., the $n$-body partition function
\be
Z_n= {(k+(1-g)(n-1))!\over n!(k-(n-1) g-1)!}
\label{lineal}\ee
indeed counts the number of ways to put $n$ particles in $k$ slots on a line with at least $g-1$ unoccupied slots
between two occupied slots, a hallmark of $g$-exclusion statistics for a lineal 1-body spectrum.
This lineal counting coincides with the original Hilbert space Haldane counting.

Note that $y_i^k$ for $i=2,$...$,g$ scale as $\sim z^{k/(g-1)} \sim e^{- \beta \mu k/(g-1)}$, and thus the $i>1$
terms in (\ref{linealbis}) make nonperturbative in $1/k$ contributions, while the extra factor with
respect to $y_1^k$ in the term $i=1$ makes perturbative in $1/k$
contributions, all becoming irrelevant in the $k\to\infty$ limit. 
Therefore, the grand partition function (\ref{linealbis}) differs from its thermodynamic limit (\ref{toto}) by both
perturbative and nonperturbative in $1/k$ terms (the corrections start at order $z^2$). This is already explicitly visible for $g=2$ statistics 
\be
{\cal Z}_2 = {y_1^{k+2} - y_2^{k+2}\over \sqrt{1+4z}}
\label{g=2Z}\ee  where the term $y_2^{k+2}$ produces nonperturbative corrections, but the
additional factor $y_1^2 /\sqrt{1+4z}$ in $y_1^{k}$ produces terms of order $k^0$, down by a factor
$1/k$ compared to the thermodynamic result (\ref{toto}). Extending the $n$-summation in (\ref{linealbis})
to infinity, including the unphysical values $n > [(k+g-1)/g]$,
 leads to  (\ref{g=2Z}) with the term $y_2^{k+2}$ absent, still
introducing perturbative corrections compared to (\ref{toto}).

We note that there exists yet another form of $g$-exclusion which admits again a valid microscopic interpretation (see again the Appendix  for details), namely
\be
{\cal Z}'_g = \sum_{i=1}^g y_i^k=\sum_{n=0}^{[k/g]} {k(k+(1-g)n-1)!\over n!(k-n g)!}z^n
\label{periodicbis}
\ee
which now corresponds to exclusion on a {\it periodic} sequence of 1-body states. Indeed,
the summation over $n$ in (\ref{periodicbis}) is truncated at $[k/g]$, as is natural for $g$-exclusion in a periodic spectrum in which $[k/g]$  particles exclude all states for additional particles. The degeneracy factor $Z'_n$ at order $z^n$, i.e., the  $n$-body partition function
\be\label{periodic}
Z'_n={k(k+(1-g)n-1)!\over n!(k-n g)!}
\ee
counts the number of ways to put $n$ particles in $k$ slots on a circle with at least $g-1$ unoccupied slots between two occupied slots, a hallmark of $g$-exclusion statistics for a 1-body periodic spectrum. 
The terms $i>1$ in (\ref{periodicbis}) make again nonperturbative in $1/k$ contributions, but the term $i=1$
is identical to the thermodynamic limit (\ref{toto}). Therefore,
the expansion in powers of $z$ of (\ref{periodicbis}) is identical to that of (\ref{toto}), but the summation over $n$ in the latter
unphysically extends beyond $[k/g]$ all the way to infinity, making a nonperturbative contribution.
This is to be  contrasted to the lineal case (\ref{linealbis}) where both nonperturbative and perturbative
corrections appear. In this sense, the periodic formulation is closer to the thermodynamic limit than the lineal one.

We conclude by pointing out that the thermodynamics of $g$-exclusion particles on an arbitrary discrete 1-body spectrum
$\epsilon_i$, $i=1,2,\dots$, and corresponding density of states $\rho(\epsilon)$ in the thermodynamic limit,
can be inferred from the physical solution $y_1$ of (\ref{1}) as
\be
\ln {\cal Z} = \sum_i \ln y_1 (z \rme^{-\beta \epsilon_i}) ~\to~ \int d\epsilon \rho(\epsilon) 
\ln y_1 (z \rme^{-\beta \epsilon})
\label{genZ}\ee
The general cluster coefficients, in particular, are obtained by using (\ref{0}) as
\be
b_n (\beta) = c_n (g) \sum_i \rme^{-n\beta \epsilon_i} ~\to~ c_n (g) \int d\epsilon \rho(\epsilon)\, \rme^{-n\beta \epsilon}
\label{genb}\ee
Note that the sum (or integral) in (\ref{genb}) is simply the 1-body partition function in the spectrum $\epsilon_i$
with temperature parameter $n\beta$. This would also arise from a path-integral representation \cite{Poly}, where
this term is the connected $n$-body path integral consisting of a single path winding $n$ times in periodic Euclidean
time, and statistics enters through the coefficients $c_n (g)$.
We stress that relations (\ref{genZ},\ref{genb})
hold in general, for any statistics that can be described in terms of an effective single-level grand partition function
and the associated  cluster coefficients.

\section{$g$-exclusion $\to (1\bb-\bb g)$-inclusion thermodynamics}\label{gto1-g}

We now turn to inclusion statistics, which we define by equations (\ref{00}-\ref{1}), as for exclusion statistics,
but now with a {\it negative} exclusion parameter. To do so, we remark \cite{usus} that turning $g\to 1-g$
in (\ref{0}) yields
\be
c_{n}(1-g)=(-1)^{n-1}c_n(g)\nonumber 
\ee
which implies for the cluster expansion (\ref{s1})
\be
\ln y(z,1-g) = \sum_{n=1}^{\infty} \,c_{n}(1-g)z^n= \sum_{n=1}^{\infty} (-1)^{n-1} \,c_{n}(g)z^n=-\ln y(-z,g)\nonumber
\ee 
where now we denote by $y(z,g)$ the  physical  single-level $g$-exclusion grand partition function satisfying
(\ref{1}). Therefore,
\be 
y(z,1\bb-\bb g)={1\over y(-z,g)}
\label{dualdual}\ee 
which provides a duality relation between $g\ge 1$ exclusion statistics and $1-g\le 0$ inclusion statistics.
This relation can also be directly derived from (\ref{1}) by rewriting it as
\be \left({1\over y(z,g)}\right)^{1-g}-\bigg({1\over y(z,g)}\bigg)^{(1-g)-1}=-z 
\nonumber
\ee
leading again  to (\ref{dualdual}). For integer $g>0$, (\ref{1}) with $g \to 1-g$ also has $g$ solutions. 
The physical inclusion solution $y_1 (z,1-g) = 1 + z + \dots=1/y_1 (-z,g)$ is recovered from the corresponding physical
exclusion solution  $y_1 (-z,g)$. 

\noindent For  $g=1\to 1-g=0$, i.e., Fermi$\;\to\;$Bose, the duality reduces
to  
\be
y_1(z,0) ={1\over 1-z} =  {1 \over y_1(-z,1)} \label{bf}
\ee
For $g=2$ exclusion , and therefore  $1-g=-1$ inclusion, the two dual solutions  are
\be
y_1(z,-1)={1\over y_1(-z,2)}={1-\sqrt{1-4z}\over 2z}~,~~~~  y_2(z,-1)={1\over y_2(-z,2)}
={1+\sqrt{1-4z}\over 2z}
\nonumber\ee
We note that  $z$ is bounded by $1/4$ and $1\le y_1 \le 2$. Similarly, for general
$(1-g)$-inclusion the bounds  become \cite{usus}
\be
0\le z\le  {(g-1)^{g-1}\over g^g} ~,~~~ 1\le y_1 \le {g \over g-1}
\nonumber\ee
The upper bounds for $z$, and especially for $y_1$, which occur only for inclusion statistics when $g<0$, are at the root of
the enhanced condensation properties
of inclusion particles and the reduction in the space dimension where condensation occurs 
(note that in  (\ref{bf}), i.e., the Fermi$\;\to\;$Bose $g=1\to 1-g=0$  case,   $z\le 1$, but $y_1$ is unrestricted).

To summarize, the duality transformation $g \to 1-g$ amounts to $y\to 1/y$ and $z\to -z$, and, in light of
(\ref{toto}), implies for a general thermodynamic grand partition function, degenerate or non-degenerate,
\be
{\cal Z}_{1-g}(z) = {1\over {\cal Z}_g(-z)}\label{dual}
\ee
We stress that the $(1-g)$-inclusion thermodynamic relations (\ref{genZ},\ref{genb}) still hold, but now in terms
of the physical solution $y (z\rme^{-\beta\epsilon},1\bb-\bb g)$ 
and the cluster coefficients $c_n(1-g)$.

\section{Microscopic exclusion statistics for a non-degenerate 1-body spectrum \label{3333}}

From here on we focus  in general  on a non-degenerate lineal  $1$-body
spectrum $\epsilon_1, \epsilon_2, \ldots, \epsilon_k$ where the number of quantum states $k$ can be finite or infinite.
The definition of  $g$-exclusion statistics  for such a  spectrum in terms of occupation numbers amounts to imposing
the constraint that individual particles are at least $g$ energy levels apart (according to a natural ordering of the
$\epsilon_i$'s).
As a consequence, the $n$-body partition function\footnote{For simplicity, above and in the rest of the paper we keep the
same notation for the grand partition functions ${\cal Z}_g$ and ${\cal Z}_{1-g}$, the ensuing $n$-body partition functions
$Z_n$, the related exclusion matrices, etc., irrespective of the 1-body spectrum being degenerate or non-degenerate.}
$Z_n$ of particles with $g$-exclusion has the form
\be
Z_{n}=\sum_{k_1=1}^{k-gn+g}\sum_{k_2=1}^{k_1}\cdots\sum_{k_n=1}^{k_{n-1}} s({k_1+gn-g})
\ldots s({k_{n-1}+g}) s(k_n)
\label{Znlin}\ee
where we defined the `spectral function' $s(i)=\exp(-\beta\epsilon_i)$. In the degenerate case at zero energy, namely  $s(1) = \cdots = s(k) = 1$, it is easy to see that $Z_n$ reduces to the lineal counting  (\ref{lineal}).  

From the $Z_n$'s in (\ref{Znlin}) we obtain the $g$-exclusion grand partition function \cite{oulabis}
\bea 
\hskip -2cm {\cal Z}_g &=& \sum_{n=0}^{\infty} Z_n z^n\nonumber\\&=&\exp\bb\bb\bigg(\bb\bb-\sum_{n=1}^{\infty}\,
(-z)^n\hskip-0.4cm
\sum_{{l_1, l_2, \ldots, l_{j}\atop \text{g-composition}\;{\rm of}\;n}}\hskip -0.3cm c_g(l_1,l_2,\dots,l_{j} )\bb
\sum_{i=1}^{k- j+1}s^{l_j}(i\bb+\bb j\bb-\bb1)\cdots s^{l_2}(i\bb+\bb1)s^{l_1}(i)\bb\bigg)~~~~\label{clark}\eea
where the $c_g(l_1,l_2,\ldots,l_{j} )$'s in the cluster expansion are known combinatorial factors, and the summation is on
$g$-compositions, which are partitions
where the ordering does matter and with up to $g-1$ consecutive zero values for $l_i$ allowed.
From (\ref{clark}) one can in turn obtain the single-level grand partition function $y(z,g)$  for  $g$-exclusion statistics in a discrete 1-body spectrum (for details on the microscopic $g$-exclusion statistics  see \cite{oulabis}).

As advocated in \cite{emeis,oulabis}, all the above information can be conveniently encapsulated in a
$(k\bb+\bb g\bb-1)$-dimensional exclusion matrix with 
a unit upper-diagonal and $g-1$ empty sub-diagonals between it and a nonzero sub-diagonal of spectral functions.
For example, for $g=3$
\bea H_3=\begin{pmatrix}
0& 1 & 0 &~~~0 & \cdots & 0  &0 &~~0  \\
0 & 0  & 1 &~~~0&\cdots  & 0&0&~~0   \\
\bb\bb -s(1) & 0 & 0 &~~~1 &\cdots &0 & 0&~~0 \\
 0&\hskip -0.2cm -s(2)&{0}&~~~0&\cdots &0 & 0&~~0\\
  \vdots&\vdots & \vdots &~~~\vdots  &{\ddots} &\vdots &\vdots &~~ \vdots\\
0 &0 & 0  &~~~0& \cdots &{0} & 1&~~0  \\
0 & 0 & 0 &~~~0&\cdots& {0}&{0}  &~~1 \\
0 & 0 & 0 &~~~0 & \cdots&\bb\bb\bb-s(k) &{0} &~~0 \\
\end{pmatrix}\label{matrixL}\eea
Its  secular determinant $\det(1- z^{1/g} H_g)$ takes the form
\be
\det(1- z^{1/g} H_g)=\sum_{n=0}^{[(k+g-1)/g]}Z_{n}\, z^{n}
={\cal Z}_g
\label{sososonice}\ee
where $Z_n$ are the $n$-body partition functions given in (\ref{Znlin}),
and thus it is  the grand partition function ${\cal Z}_g$ of a gas of particles with $g$-exclusion in
the 1-body lineal
spectrum $\epsilon_1, \epsilon_2, \ldots, \epsilon_k$.

We mention that a similar matrix formulation is also available for  a periodic spectrum (where 1-particle states $i$
are positioned on a circle and $\epsilon_k$ and $\epsilon_1$ are neighbors), involving a periodic version of the
lineal exclusion matrix above and reproducing  
the periodic counting (\ref{periodic}) in the degenerate case up to some adjustments (see the Appendix  for details). 
In the rest of the paper we will focus on the case of a lineal 1-body spectrum, relegating the case
of a periodic 1-body spectrum, in which some additional subtleties arise, to the Appendix. 

\section{Microscopic inclusion statistics for a non-degenerate 1-body spectrum \label{yesitis}}

For inclusion statistics we do not have an {\it a priori} microscopic definition, such as the ``minimal distance $g$" rule
between filled levels in exclusion statistics. We have to devise such a formulation, on the basis of some
assumptions, and give its combinatorial interpretation. The requirements are that it must
involve integer and nonnegative multiplicities of microscopic states, and must also lead to the proper
thermodynamic limit of section \ref{gto1-g} for a degenerate spectrum when the number of 1-body quantum states
$k$ becomes infinite.

We recall that for non-interacting particles, denoting by $n_i$ the number of particles occupying the energy level
$\epsilon_i$ (i.e., the occupation number of $\epsilon_i$), the $n$-body energy is expressed as the simple sum
$E_n = \sum_{i=1}^{\infty} n_i\epsilon_i$  with $n = \sum_{i=1}^{\infty} n_i$.  In the Bose and Fermi cases, where the occupation numbers $n_i$ are independent of each other and the multiplicity of a $n$-body state with
a given $n_1, n_2, \ldots, n_k$ is trivially equal to 1, the grand partition function
 \bea 
\quad {\cal Z} = \sum_{n=0,E_n}^{\infty} z^n \rme^{-\beta E_n} \nonumber= \sum_{n_1,...,n_k} \big(z \rme^{-\beta \epsilon_1}\big)^{n_1}\ldots\big(z \rme^{-\beta \epsilon_k}\big)^{n_k}
= \sum_{n_1} \big(z \rme^{-\beta  \epsilon_1}\big)^{n_1}\ldots\sum_{n_k} \big(z \rme^{-\beta  \epsilon_k}\big)^{n_k}\eea
rewrites in the Bose case, where $n_i = 0,1,...,\infty$,  as
\be\nonumber {\cal Z}_0 = \Big({1\over 1-z\rme^{-\beta \epsilon_1}}\Big)\ldots\Big({1\over 1-z\rme^{-\beta \epsilon_k}}\Big)\ee
and in  the Fermi case, where $n_i = 0,1$, as
\be {\cal Z}_1= \big(1+z\rme^{-\beta \epsilon_1}\big)\ldots\big(1+z\rme^{-\beta \epsilon_k}\big)\nonumber\ee
so that the standard microscopic Fermi-Bose correspondence
\be\nonumber {\cal Z}_0(z)={1\over {\cal Z}_1(-z)}\ee 
 appears here as the $g=1\to 1-g=0$ case of the thermodynamic duality relation (\ref{dual}).

To derive a microscopic definition of inclusion statistics, we start from  (\ref{dual})
and postulate that it is also valid  for the microscopic grand partition functions. This implies,
in terms of  the inverse of the secular determinant \be\nonumber
{\cal Z}_{1-g}={1\over {\cal Z}_{g}|_{z\to -z}}={1\over \det(1-  z^{1/g} H_g)|_{z\to -z}}
\ee
(see \cite{math} for studies of the inverse of characteristic polynomials
of matrices from a combinatorial perspective).
For $g=2$, i.e., inclusion $1-g=-1$, this yields 
\be
{\cal Z}_{-1}=\bb\sum_{n_1,\ldots,n_k=0}^{\infty}\bb (zs(1))^{n_1}{n_1+n_2\choose n_1}
(zs(2))^{n_2} {n_2+n_3\choose n_2}\ldots
{n_{k-1}+n_k\choose n_{k-1}} (zs(k))^{n_k} 
\nonumber\ee
while for $g=3$, i.e., inclusion $1-g=-2$,
\bea 
{\cal Z}_{-2}= 
\bb\sum_{n_1,\ldots,n_k=0}^{\infty}\bb &&(zs(1))^{n_1}{n_1+n_2+n_3\choose n_1}(zs(2))^{n_2}{n_2+n_3+n_4\choose n_2}\ldots
\nonumber\\&&\ldots{n_{k-2}+n_{k-1}+n_k\choose n_{k-2}}(zs(k-1))^{n_{k-1}}{n_{k-1}+n_k\choose n_{k-1}}(zs(k))^{n_k}\nonumber\eea
etc. This leads to nontrivial inclusion multiplicities {$m_{1-g} (n_1,\dots,n_k )$ for states of given
occupation numbers $n_1,\dots,n_k$:}
for $g=2\to 1-g=-1$ they are 
\be\nonumber
{ m_{-1} (n_1,\dots,n_k) =} {n_1+n_2\choose n_1}{n_2+n_3\choose n_2}\ldots{n_{k-1}+n_k\choose n_{k-1}}
\ee
and are indeed related to the degenerate  grand partition function  (\ref{linealbis}), that is, they yield  as they should the
 inverse of ${\cal Z}_2$ given in (\ref{g=2Z}) with $z\to -z$
\be
{\cal Z}_{-1}={\sum_{n_1,\ldots,n_k=0}^{\infty}  m_{-1} (n_1,\dots,n_k)\, z^{n_1+n_2+\ldots +n_k} }
= {\sqrt{1-4z} \over y^{k+2}_1(-z,2) - y_2^{k+2}(-z,2)} 
\label{sibien}\ee
For $g=3\to 1-g=-2$ inclusion they are
\be
{ m_{-2} (n_1 ,\dots,n_k ) =} {n_1+n_2+n_3\choose n_1}{n_2+n_3+n_4\choose n_2}\ldots
{n_{k-2}+n_{k-1}+n_k\choose n_{k-2}}{n_{k-1}+n_k\choose n_{k-1}}
\nonumber\ee
with
${\cal Z}_{-2}=\sum_{n_1,\ldots,n_k=0}^{\infty} m_{-2} (n_1 ,\dots,n_k )\, z^{n_1+n_2+\ldots +n_k}$
 the corresponding degenerate grand partition function obtained by inverting (\ref{linealbis}) in the case
$g=3$ and trading $z\to -z$.

Generalizing the multiplicities to $(1-g)$-inclusion is straightforward. The resulting multiplicities admit the
interpretation of
the product of multiplicities of clusters of $g$ adjacent states with the particles in each cluster considered
as distinguishable, divided by the corresponding multiplicities of overlaps between clusters. Defining the
shorthand notation
\be
[i,j] := {{n_i + n_{i+1} + \cdots + n_j} \choose n_i \,,\, n_{i+1}\, ,\, \dots\, ,\, n_j}
\nonumber\ee
 the multiplicity\footnote {There is a formula analogous to (\ref{lineal}) for the $(1-g)$-inclusion degenerate lineal counting
 \bea Z_n&=&\sum_{n_1,\ldots,n_k=0}^{\infty}  m_{1-g} (n_1 ,\dots, n_k )\; \delta_{n_1+\ldots+n_k,n}\nonumber\eea
\vskip -0.1cm \noindent
valid for $n\le k/(g-1)+1$, namely
\bea Z_n&=&{((k+g)(k+g-1)-g(g-1)n)(k+g+gn-2)!\over n!(k+g+(g-1)n)!}\nonumber\eea
which, for example in the case $g=2\to 1-g=-1$, leads to the generating function
\be\nonumber \sum_{n=0}^{\infty}Z_n z^n=\sqrt{1-4z}/\;y_1^{k+2}(-z,2)\;\nonumber\ee
\vskip -0.1cm\noindent
i.e., (\ref{sibien}),  which is the inverse of ${\cal Z}_2 $ in (\ref{g=2Z}) with $z\to -z$ and only the term $y_1$ kept. Likewise for $g=3$ we would obtain the generating function \be\nonumber \sum_{n=0}^{\infty}Z_n z^n=(y_1(-z,3)-y_2(-z,3))(y_1(-z,3)-y_3(-z,3))/\;y_1^{k+4}(-z,3)\;\nonumber\ee   where the $y_i$ are given in (\ref{the3}), again the inverse of (\ref{linealbis}) for $g=3$ with $z \to -z$ and only the term $y_1$ kept. The terms $n>k/(g-1)+1$ make nonperturbative contributions.} for $k\ge g$ is
\be
{ m_{1-g} (n_1 ,\dots, n_k ) =\;} {[1,g]\, [2,g+1] \cdots [k+1-g,k] \over [2,g]\, [3,g+1] \cdots [k+1-g,k-1]}
\nonumber\ee
while for $k < g$ there is a single cluster, and the  multiplicity reduces to
\be
[1,k] = {{n_1 + n_2 + \cdots + n_k}\choose {n_1 ,\, n_1 ,\, \dots,\, n_k}} \,,
\nonumber\ee
i.e., the one for distinguishable particles.
 Its generating function 
\be
\sum_{n_1, \dots,n_k=0}^{\infty} [1,k]\, z_1^{n_1} \cdots z_k^{n_k} = {1 \over 1-z_1 - \cdots -z_k}
= \sum_{n=0}^\infty (z_1 + \cdots + z_k )^n
\nonumber\ee
is the grand partition function of distinguishable particles on the $k$ levels $\epsilon_1, \epsilon_2, \ldots, \epsilon_k$, with $z_i = z e^{-\beta \epsilon_i}$. So the limit $g\to
\infty$ that is to say $1-g \ll -1$ corresponds to distinguishable statistics.
However, the thermodynamic limit is not necessarily that of distinguishable particles, as the limits $k \to \infty$
and $g \to \infty$ do not commute, and the result depends on their relative scaling.

\noindent These multiplicities generalize the Bose and Fermi multiplicities, trivially equal to $1$ (or $0$ in the
Fermi case if any $n_i >1$), by enhancing the weights of neighboring clusters of occupied $1$-body levels.
This enhancement
is the counterpart of the dilution of particles in exclusion statistics and a hallmark of inclusion statistics.

\section{Inclusion statistics in a harmonic well}

As a concrete  example let us consider particles with $g$-statistics in a one-dimensional harmonic well, populating the equidistant  $1$-body spectrum
$\epsilon_i={ i}\omega$, $i=0,1,2,\dots, \infty$, with spectral function $s(i)=x^{i}$ (we denote $x=\rme^{-\beta\omega}$).
Since there is
no upper bound in the energy, this corresponds to a semi-infinite spectrum where $k=\infty$.

\subsection{$g$-exclusion}

For $g\ge 0$, as is well known \cite{Poly0,Das}, $g$-exclusion statistics is realized by the $n$-body Calogero Hamiltonian \cite{Calogero}
\be\label{caloH}
H_n= -\frac{1}{2}\sum_{i=1}^n \frac{\partial^2}{\partial x_i^2} + 
\sum_{i<j}\frac{g(g-1)}{(x_i-x_j)^2}
+\frac{1}{2}\,\omega^2\, \sum_{i=1}^n x_i^2
\ee
with $n$-body spectrum
\be
E_n= \omega\Big(\sum_{i=1}^n l_i +g{n(n-1)\over 2}  \Big)\, ,\quad 0\le l_1\le l_2\le \ldots\le l_n
\label{caloS}\ee
(we eliminated the trivial zero-point energy per particle $\omega/2$ to make the spectrum conform with the
convention $\epsilon_i = i \omega$, $i=0,1,2,\dots$). This can be rewritten
in terms of the quasi-excitation numbers $l'_i=l_i+g(i-1)$ as
\be
E_n= \omega\sum_{i=1}^n l'_i \quad   0\le l'_i\le l'_{i+1}-g,\ldots, 0\le l'_n\le \infty
\label{000}\ee
where the $l'_i$ are indeed separated by at least $g-1$ energy quanta, which is nothing but  $g$-exclusion.
From this $n$-body  spectrum we get the partition function $Z_n$
\bea
Z_n\nonumber &=&x^{g n (n - 1)/2}\bb\bb\bb\bb\sum_{\quad 0\le l_1\le l_2\le \ldots\le l_n\le \infty}x^{\sum_{i=1}^n l_i} \\
&=&\bb\bb\bb\bb\sum_{\quad   0\le l'_i\le l'_{i+1}-g,\ldots, 0\le l'_n\le \infty}x^{\sum_{i=1}^n l'_i}
\nonumber\\
&=&~x^{g n (n - 1)/2} \prod_{j=1}^n{1 \over 1 - x^j}
\label{cestcabis}\eea
We stress here that the $n$-body partition function obtained above from the $n$-body Calogero spectrum could
as well be derived from the very definition of $g$-exclusion statistics by means of the $Z_n$'s given in (\ref{Znlin}),
in this case for the spectral function $s(i)=x^{i}$ and the number of 1-body quantum states  $k\to\infty$. This
is reconfirming that the $n$-body Calogero Hamiltonian  (\ref{caloH}) is a microscopic dynamical realization of
$g$-exclusion statistics. This  leads to the grand partition function and cluster expansion \cite{oulabis}
\be
{\cal Z}_g =\sum_{n=0}^{\infty} Z_n z^n 
=\exp\bb\bigg(\bb\bb-\sum_{n=1}^{\infty}\;(-z)^n {1\over 1- x^n}\bb\bb\sum_{{l_1, l_2, \ldots, l_{j}\atop \text{g-composition}\;{\rm of}\;n}}\bb\bb c_g(l_1,l_2,\ldots,l_{j} ) \; x^{\sum_{i=1}^j(i-1)l_i}\bb\bigg)
\label{cici}\ee
where by substituting $s(i)=x^i$ and $k\to\infty$ in (\ref{clark}) we obtained in (\ref{cici})  the $g$-exclusion grand partition function for a  1-body harmonic spectrum.

\noindent
We note that the Hamiltonian (\ref{caloH}) is invariant under the mapping $g \to 1-g$. However, the spectrum
(\ref{caloS}) is not, since it depends linearly on $g$. The hermiticity properties of the Hamiltonian (\ref{caloH})
restrict the allowed wavefunctions and impose
choosing the greater of $g,1\bb-\bb g$ {{in (\ref{caloS})}}, leading naturally to exclusion statistics. This is also an early
indication that the transition to inclusion statistics will not be trivially obtained by simply trading $g\bb \to\bb 1\bb-\bb g$ in (\ref{caloS}).

\subsection{$(1\bb-\bb g)$-inclusion}

\noindent To reach $(1\bb-\bb g)$-inclusion statistics we use the duality relation (\ref{dual}) , inverting the grand partition function ${\cal Z}_g$ and turning $z\to -z$, or equivalently multiplying the $n^{th}$ order cluster coefficient by 
$(-1)^{n-1}$. This amounts to considering the cluster expansion (\ref{cici}) with the minus signs removed, namely
\bea 
{\cal Z}_{1-g} &=& \exp\bigg(\sum_{n=1}^{\infty} {z^n\over 1- x^n}\sum_{{l_1, l_2, \ldots, l_{j}\atop \text{g-composition}\;{\rm of}\;n}}c_g(l_1,l_2,\ldots,l_{j} ) \; x^{\sum_{i=1}^j(i-1)l_i}\bigg)
\nonumber\eea
which in turn can be expanded in $z$
\bea\nonumber
{\cal Z}_{1-g} &=&\sum_{n=0}^{\infty} Z_n z^n
\eea
leading to the new $n$-body partition function\footnote{It rewrites equivalently as 
\bea \nonumber Z_n&=&\bb\bb
\sum_{-(g-1) \big( (n - 1)/2 -(i-1)\big) \le l_i\le l_{i+1},\ldots, (g-1) (n - 1)/2 \le l_n\le \infty}x^{\sum_{i=1}^n l_i}\nonumber\\&=&  x^{(1-g) n (n - 1)/2}\nonumber
\sum_{(g-1)  (n-1)/2\le l_i\le l_{i+1}+(g-1),\ldots,(g-1) (n-1)/2\le l_n\le \infty}x^{\sum_{i=1}^n l_i}\nonumber\eea}
\bea\nonumber
Z_n&=&  x^{(1-g) n (n - 1)/2}\sum_{\quad  (g-1)(i-1)\le l_i\le l_{i+1},  \ldots, (g-1)(n-1)\le l_n\le \infty}x^{\sum_{i=1}^n l_i}\\
&=&\bb\sum_{\quad   0\le l'_i\le l'_{i+1}-(1-g),\ldots, 0\le l'_n\le \infty}x^{\sum_{i=1}^n l'_i}
\label{01}\eea
where $l'_i = l_i +(1-g)(i-1)$ are a new set of quasi-excitation numbers.
One can check that the ($1-g$)-inclusion partition functions   (\ref{01}) are indeed identical to those derived from the inclusion
occupation multiplicities introduced in Section (\ref{yesitis}).
For example, the $g=2\to1-g=-1$ inclusion 2-body and 3-body partition functions are  from (\ref{01}) 
\be 
Z_2= {1 + x - x^2\over (1 - x) (1 - x^2)}~,\quad\quad
Z_3= {1 + 2 x - x^3 - 2 x^4 + x^6\over(1 - x) (1 - x^2) (1 - x^3)}
\label{ci} \ee
In the 2-body case, the multiplicities are $2$ for adjacent particles ($n_i=n_{i+1}=1$) and $1$
for all other configurations. This yields the 2-body partition function
\be
\sum_{k=0}^{\infty} x^{2k}(1+2x+x^2+x^3+\ldots)={1\over 1-x^2}\left({1\over 1-x}+x\right)
\nonumber\ee
which is identical to $Z_2$ in (\ref{ci}).
Similarly, in the 3-body case one obtains  starting from the configurations with at least one particle populating  the 1-body ground state, then no particle in the ground state but at least one  populating  the first excited state etc., and plugging the relevant multiplicities
\bea 
&&\sum_{k=0}^{\infty} x^{3k}(1 + 3  x + 4 x^2 + 5 x^3 + 4 x^4 + 5 x^5 + 5 x^6 + 6 x^7 + 6 x^8 + 
 7 x^9 + 7 x^{10} +\ldots)\nonumber\eea
which reproduces  $Z_3$ in (\ref{ci}). 

Note that the 2-body and 3-body
partition functions  (\ref{ci}) become the bosonic ones when one replaces their numerators by 1. The polynomials in the numerators, then, account for the additional degeneracies introduced by inclusion statistics.
This is entirely general: the $n$-body partition function for $(1\bb-\bb g)$-inclusion is given by
\be
Z_n = {P_{g,n} (x) \over  \prod_{j=1}^n (1 - x^j )}
\nonumber\ee
where $P_{g,n} (x)$ is a polynomial of degree $gn(n-1)/2$ that satisfies $P_{g,n} (0) = P_{g,n} (1) =1$.
The relation $P_{g,n} (0) =1$ expresses the non-degenerate nature of the ground state,
while $P_{g,n} (1) =1$ arises from the classical thermodynamic limit: for $x \to 1$ ($\omega \to 0$) the $n$-body
partition function becomes independent of statistics, and thus $P_{g,n} (1)$ for inclusion and the corresponding
factor $x^{gn(n-1)/2}$ in (\ref{cestcabis}) for exclusion, must become $1$. The qualitative difference between
$P_{g,n} (x)$ and $x^{gn(n-1)/2}$ demonstrates again
the nontrivial nature of the transition from $g$-exclusion to $(1\bb-\bb g)$-inclusion.

The harmonic trap potential serves as a ``box" confining the particles. Reducing its strength
enlarges the box and amounts to increasing the one-dimensional volume. Therefore, the thermodynamic limit
$x\to 1$ ($\omega\to 0$) at constant chemical potential means the infinite volume limit, {and we can
apply the results of \cite{usus} to recover the thermodynamic properties of the system. The 1-body spectrum becomes dense
in this limit, with a constant density of states $\rho(\epsilon) = 1/\omega$ typical of a free $2$-dimensional system, and therefore corresponds to the $2$-dimensional
case in \cite{usus} upon identifying the  volume (area) with $h^2/(2\pi m \omega)$. This results in the
 gas of particles in a harmonic trap undergoing condensation for {\it any} inclusion $1\bb-\bb g <0$ at critical temperature
\be
T_c = {\omega N \over k \ln{g\over g-1}}
\ee
with $N$ the number of particles  and $k$ Boltzmann's constant. This is to be contrasted to bosons, which
do not condense in a 1-dimensional harmonic  trap potential (whereas they do condense in a 2-dimensional harmonic trap potential).

It should be stressed that the $\omega \to 0$ limit is distinct from the infinite-volume limit in free space, which leads to
the density of states $\rho(\epsilon) \sim \epsilon^{-1/2}$. The free infinite-volume limit can nevertheless be recovered from the
$\omega\to0$ limit by scaling $z$ such that the particle density around the origin remains finite and extracting the
corresponding part of the (extensive) grand potential. For details on this thermodynamic limit procedure see \cite{Das} and
the Appendix.}

We note that  from  the $Z_n$'s in (\ref{01}) we can extract the surprisingly simple $(1-g)$-inclusion statistics Calogero-like spectrum  
\be
E_n= \omega\Big(\sum_{i=1}^n l_i +(1-g){n(n-1)\over 2} \Big)\, ,\quad  
(g-1)(i-1)\le l_i\le l_{i+1}, ~ (g-1)(n-1)\le l_n
\label{0001}\ee
or equivalently, in terms of the quasi-excitations $l'_i$,
\be
E_n= \omega\sum_{i=1}^n l'_i \, ,\quad  0\le l'_i\le l'_{i+1}+g-1 \, , ~ 0\le l'_n
\label{0000}\ee
We  insist  that the na\"ive expectation
that the inclusion spectrum (\ref{0000}) is obtained from the spectrum
(\ref{000}) by simply substituting $g \to 1-g$ is incorrect. This is due to the inequalities
$ (g-1)(i-1)\le l_i$ in (\ref{0001}), or equivalently
$0\le l'_i $ in (\ref{0000}): the corresponding $l'_i$ defined by simply taking the exclusion $l'_i = l_i +g(i-1)$
in (\ref{000}) and turning $g \to 1-g$
satisfy $l'_i \ge (1-g)(i-1)$ and thus can dip below zero for $i>1$. The two spectra differ by the states produced
by such negative values of $l'_i$. In particular, the ground state energy of the na\"ive spectrum (\ref{000})
with $g \to 1-g$ is negative, while the ground state of (\ref{0000}) is at zero energy and nondegenerate.
The na\"ive spectrum obtained by substituting $g \to 1-g$  in (\ref{caloS}),
namely
 \be
E_n= \omega\Big((1-g){n(n-1)\over 2} +\sum_{i=1}^n l_i \Big) ,\quad 0\le l_1\le l_2\le \ldots\le l_n
\nonumber\ee
gives rise to the $n$-body partition function
\be
Z_n = x^{(1-g) n (n - 1)/2}\hskip -0.4cm \sum_{\quad 0\le l_1\le l_2\le \ldots\le l_n\le \infty}\bb
x^{\sum_{i=1}^n l_i} = {x^{(1-g)n(n-1)/2} \over \prod_{j=1}^n (1 - x^j )}\nonumber
\ee
The thermodynamic limit of this model, obtained from ${\cal Z}_g = \sum_n Z_n z^n$ with $Z_n$ as above
and $x\to 1$, $z \to 0$, does not exhibit condensation and does not reproduce the thermodynamics of the inclusion model.

\section{{\bf Conclusions}}
We obtained a microscopic description of inclusion statistics in terms of the inverse of the grand partition
function of exclusion statistics, and re-expressed it in terms of nontrivial occupation
number multiplicities that enhance the weights of neighboring clusters of 1-body quantum states, a key
property of inclusion statistics. We also presented a duality relation that maps $g$-exclusion to 
$(1\bb\bb-\bb\bb g)$-inclusion statistics, the well-kown Bose-Fermi
correspondence appearing as the $g=0$ special case.
 
The expressions of the aforementioned inclusion multiplicities are relatively simple and intuitive, a fact not a priori obvious,
since they are obtained by inverting an already quite non-trivial expression for the $g$-exclusion grand partition function. 
The same can be said of the quite simple $n$-body Calogero-like spectrum that arises from the inversion of the Calogero
grand partition function and is paramount to inclusion statistics in a harmonic 1-body spectrum.
 
There are several open questions or directions for further investigation. On the mathematics side, the
combinatorial properties of inclusion statistics appear to be quite rich and deserve further study.
In addition, in view of the known connection of exclusion statistics and planar lattice paths \cite{emeis} or
forward-moving paths of Dyck, Motzkin, and Lukasiewicz type \cite{em1,em2,em3}, it is worth investigating the
potential connection of inclusion statistics with particular types of lattice path or other processes.

The most interesting remaining questions are those related to the physics of inclusion statistics.
As already stressed, the experimental consequences of inclusion are striking, in particular due to the propensity
of inclusion particles to achieve macroscopic condensation in planar systems. Therefore, the realization of inclusion
statistics in a concrete experimental situation is of physical relevance. 

Finally, the realization of the
Calogero-like $n$-body spectrum for $(1-g)$-inclusion statistics in (\ref{0000}) in terms of an
$n$-body Calogero-like Hamiltonian remains a fascinating open issue.

\noindent {\bf Acknowledgements}

\noindent A.P.'s research was supported by NSF under grant NSF-PHY-2112729 and by PSC-CUNY under grants 65109-0053 and 6D136-0004. A.P. also acknowledges the support of a PALM grant and the hospitality of LPTMS, CNRS at Université Paris-Saclay (Faculté des Sciences d’Orsay), where this work was initiated. \\
\noindent S.O. would like to thank Li Gan for technical help and a careful reading of the manuscript. S.O. also acknowledges the hospitality of the City College of New York,
 where part of this work was done.

\section{{\bf Appendix}}
\subsection{Periodic spectrum and periodic multiplicities }

The circular periodic counting (\ref{periodic}) for a periodic degenerate $1$-body spectrum can also be formulated more generally for a  periodic nondegenerate spectrum $\epsilon_1,  \epsilon_2, \dots, \epsilon_k$. Again the key
ingredient is the $n$-body partition function for particles with $g$-exclusion in the above 1-body periodic spectrum
\be\label{perio}
Z'_{n}=\sum_{k_1=1}^{k-gn+\min(g,k_n )}\sum_{k_2=1}^{k_1}\cdots\sum_{k_n=1}^{k_{n-1}} s({k_1+g(n-1)})
\ldots s({k_{n-1}+g}) s(k_n)
\ee
Clearly (\ref{perio}) reduces to the $g=2$ periodic counting (\ref{periodic}) when the
spectrum  is degenerate. Note that (\ref{perio}) requires $k \ge g n$. For $k < g$, even a single particle excludes
itself through its periodic image and $Z_1 =0$.
The corresponding  grand partition function is
\be 
{\cal Z}'_g =\sum_{n=0}^{[k/g]} Z'_n z^n
\nonumber\ee

All this information can be encapsulated in  the $k$-dimensional 
periodic  $g$-exclusion matrix, for $k>g$, which is similar to the lineal one in (\ref{matrixL}) but with the off-diagonals
wrapping around the matrix. For example, for $g=2$, 
\be H'_2=\begin{pmatrix}
\;0\; & \;1\; & \;0\; & \cdots & 0 & -s(k) \\
-s(1) & 0 & 1 & \cdots & 0 & 0 \\
\vdots & \vdots & \vdots & \ddots & \vdots & \vdots \\
0 & 0 & 0 & \cdots &0 & 1 \\
1 & 0 & 0 & \cdots & -s(k-1) & 0\\
\end{pmatrix}\nonumber\ee
Its secular determinant reproduces ${\cal Z}'_2$ up to a residual term,
\be
\det(1- z^{1/2} H'_2)=
\sum_{n=0}^{[k/2]}Z'_{n} z^{n}+\Big(\bb-1-\prod_{i=1}^k \big(-s(i)\big)\Big)z^{k/2}
\nonumber\ee
where $\big(\bb-\bb 1\bb-\prod_{i=1}^k (-s(i))\big) z^{k/2}$ is a spurious ``Wilson loop" contribution that can be
consistently discarded. Similar results hold for $g>2$, where the terms $-s(i)$ in the $k$-dimensional matrix
$H'_g$ ($k>g$) are in the ($g-1$)-lower diagonal wrapping periodically around the matrix. In that case,
\be
\det(1- z^{1/g} H'_g)=
\sum_{n=0}^{[k/g]}Z'_n z^n +W_g (z)
\nonumber\ee
where now the spurious term $W_g$ starts at order $\sim z^{k/\big(g(g-1)\big)}$ and goes up to $z^{k/g}$.
For $g=2$, this term can be eliminated by an appropriate modification of the matrix that slightly complicates its form,
but its elimination for $g>2$ becomes more involved.

Microscopic $(1-g)$-inclusion statistics on a periodic 1-body spectrum must be defined such that it satifies appropriate
physical and consistency conditions: it must give the correct thermodynamic limit and involve integer, non-negative
state multiplicities. Further, the multiplicities must locally agree with these for lineal counting, meaning that the
state multiplicity of a cluster of particles on 1-body states that span a small subset of the full spectrum must be
the same as those for the corresponding cluster on a lineal spectrum, otherwise they could ``sense" the topology
of the spectrum, introducing a nonlocal element. These conditions largely fix the definition of periodic microscopic
inclusion statistics, but still leave some distinct possibilities.

One approach is to postulate an exact duality with $g$-exclusion statistics, and define the $(1-g)$-inclusion grand
partition function as the inverse of the periodic microscopic $g$-exclusion partition function with $z \to -z$, as in the
lineal case. Taking the $g=2$ case as an example, the grand partition function on $k$ degenerate periodic  levels is
given in (\ref{periodicbis}) as ${\cal Z}_2=y_1 (z,2)^k + y_2(z,2)^k$. Thus the resulting $(1-g=-1)$ inclusion
grand partition function obtained by duality is
\bea  
{\cal Z}'_{-1} = {1\over y^k_1(-z,2)+y^k_2(-z,2)}= {1\over  y_1^{-k}(z,-1)+y_2^{-k} (z,-1)} 
\nonumber\label{etnon}\eea
and for general $g \to 1-g$ and a degenerate spectrum,
\be
{1\over {\cal Z}'_{1-g}} = \sum_{i=1}^g { y_i^k (-z,g)}=\sum_{i=1}^g { y_i^{-k} (z,1-g)}
\label{inv2c}\ee
This definition fulfills all the desired conditions.

A different approach to periodic microscopic inclusion would be to define the occupation number multiplicities
by periodizing the lineal multiplicities of Section (\ref{yesitis}) in an obvious way.
For the simplest case of $g=2 \to 1-g=-1$ one would write
\be
m'_{-1} (n_1,\dots,n_k ) = {n_1+n_2\choose n_1}{n_2+n_3\choose n_2}\ldots
{n_{k-1}+n_k\choose n_{k-1}}{n_{k}+n_1\choose n_{k}}
\nonumber\ee
Likewise, for $g=3\to 1-g=-2$,
\be
m'_{-2} (n_1,\dots,n_k ) =
{n_1\bb+\bb n_2\bb+\bb n_3\choose n_1}{n_2\bb+\bb n_3\bb+\bb n_4\choose n_2}\ldots
{n_{k-1}\bb+\bb n_k\bb+\bb n_1\choose n_{k-1}}{n_k\bb+\bb n_1+n_2\choose n_{k}}
\nonumber\ee
and in general, for $(1-g)$-inclusion,
\be
m'_{1-g} (n_1,\dots,n_k ) = {[1,g]\, [2,1+g] \cdots [k,k+g-1] \over [1,g-1]\, [2,g] \cdots [k,k+g]}
\nonumber\ee
From the above multiplicities one can obtain the grand partition function for a degenerate spectrum.
For $g=2\to 1-g=-1$ we obtain
\bea
\hskip -0.5cm{\cal Z}'_{-1} =\sum_{n_1,...,n_k=0}^{\infty}\bb m'_{-1} (n_1,\bb...,n_k ) z^{n_1+\cdots+ n_k}
&=& {1\over  y_1^{k}(-z,2)-y_2^{k} (-z,2)}\cr &=&{1\over  y_1^{-k}(z,-1)-y_2^{-k} (z,-1)}
\label{period2c}\eea
Comparing with (\ref{inv2c}), we see that it differs only in the sign of the term $y_2$, and thus the two formulae
differ only by nonperturbative in $1/k$ terms. This shows that both definitions, via duality of by periodizing the
$m_{1-g}$, lead to the same multiplicities for a number of particles $n<k/g$, which are locally the same as the
corresponding lineal multiplicities, but they start diverging as the particles populate the full 1-body spectum.

We conclude with the remark that one could calculate a periodic inclusion grand partition function for a degenerate
spectrum by simply substituting $g\to 1-g$ in the periodic counting formula (\ref{periodic}). Unlike in the $g$-exclusion
case, the corresponding $1-g$ counting formula gives positive and integer results for all $n$, including $n>k/g$.
This approach is less useful than the previous ones, as it does not obviously generalize to a nondegenerate
spectrum, but nevertheless leads to
\be
{\cal Z}'_{1-g} =\sum_{n=0}^{\infty} {k(k+gn-1)!\over n!(k+gn-n)!} z^n = y_1^{-k}(-z,g)= y_1^k (z,1-g)
\nonumber\ee
a simple result that, again, differs from (\ref{inv2c}) and (\ref{period2c}) only in nonperturbative terms.

\subsection{Proof of the generating function formulae (\ref{linealbis}) and (\ref{periodicbis})}

Consider the generating function ${\cal Z}_{g,k}$ of $g$-exclusion particles on $k$ degenerate 1-body states
on a line, and the corrsponding one ${\cal Z}'_{g,k}$ on a circle (resp. lineal and periodic counting).
For ${\cal Z}_{g,k}$, focusing on the state at the end of the chain and putting it to be either empty, leaving
an unrestricted system on $k-1$ levels, or occupied, excluding $g$ levels and leaving an unrestricted
system on $k-g$ levels, we have
\be
{\cal Z}_{g,k} = {\cal Z}_{g,k-1} + z {\cal Z}_{g,k-g}
\label{Zrec}\ee
For ${\cal Z}'_{g,k}$, we focus on a set of $g-1$ adjacent states. There can be either no particles in these
states, leaving an open (lineal) chain with $k-g$ states, or 1 particle, excluding $g-1$ states on each side and
leaving an open chain with $k-2(g-1)-1$ states. We obtain
\be
{\cal Z}'_{g,k}= {\cal Z}_{g,k-g+1} + (g-1) z {\cal Z}_{g,k-2g+1}
\label{Ztrec}\ee
Applying (\ref{Zrec}) in (\ref{Ztrec}) we obtain
\bea
{\cal Z}'_{g,k}&=& {\cal Z}_{g,k-g} + z {\cal Z}_{g,k-2g+1} + (g-1) z \left({\cal Z}_{g,k-2g}+ z {\cal Z}_{g,k-3g+1}\right)\cr
&=& {\cal Z}_{g,k-g} + (g-1) z {\cal Z}_{g,k-2g} + z\left[ {\cal Z}_{g,k-2g+1} + (g-1) z {\cal Z}_{g,k-3g+1}\right]\cr
&=& {\cal Z}'_{g,k-1} + z {\cal Z}'_{g,k-g}\nonumber
\eea
(Note that the same argument works by focusing on $g$ adjacent states, but for no other number.
Had we chosen fewer adjacent states, the ends of the broken open chain would be subject to exclusion
constraints and would not reproduce ${\cal Z}$; had we chosen more states, we could have placed more than one
particle in them.)

Thus, both ${\cal Z}_{g,k}$ and ${\cal Z}'_{g,k}$ satisfy the same linear recursion relation in $k$.
The solution of this recursion equation can be obtained in the usual way by taking an exponential ansatz
$y^k$, which must satisfy the characteristic equation (\ref{1}), that is, the exclusion equation.
The general solution will be
\be
A_1 y_1^k + \cdots + A_g y_g^k
\nonumber\ee
with $y_i$ the $g$ solutions of (\ref{1}) and $A_i$ arbitrary coefficients.

${\cal Z}_{g,k}$ and ${\cal Z}'_{g,k}$ differ only in their initial conditions. We have
\bea
&&{\cal Z}_{g,k} = 1+ k z ~,~~ k = 1,2,\dots,g 
\label{iniZ}
\\
&&{\cal Z}'_{g,k}= 1 ~,~~ k=1,\dots,g-1 ~;~~~{\cal Z}'_{g,g} = 1+ g z \label{iniZt}\eea
since there can be either no particles (on the periodic spectrum) or 1 particle (on the lineal spectrum) for $k<g$,
and 1 particle for $k=g$.
Writing (\ref{1}) in factorized form in terms of its roots $y_i$ and taking its logarithm gives
\be
\sum_{i=1}^g \ln (y - y_i) = \ln (y^g - y^{g-1} -z )
\nonumber\ee
Expanding the two sides in $1/y$ and matching powers yields
\bea \nonumber
&& \sum_{i=1}^g y_i^k = 1 ~,~~ k=1,2,\dots,g-1 
\\
&& \sum_{i=1}^g y_i^g = 1+ g z \nonumber
\eea
Comparing this with (\ref{iniZt}) we see that $A_1 = \cdots = A_g = 1$ for ${\cal Z}'_{g,k}$, so
\be
{\cal Z}'_{g,k}= \sum_{i=1}^g y_i^k\nonumber
\ee
i.e., (\ref{periodicbis}). Note that $y_i^k$ for $i>1$ make purely nonperturbative contributions.

For ${\cal Z}_{g,k}$ we need to solve the $g$ linear equations (\ref{iniZ}) for the $A_i$ in terms of the
$y_i$. This is tedious, but the result can be obtained analytically.
For $g=2$ we find
\be
{\cal Z}_{2,k} = {y_1^{k+2} - y_2^{k+2} \over y_1 - y_2} = {y_1^{k+2} - y_2^{k+2}\over\sqrt{1+4z}}\nonumber
\ee
for $g=3$
\bea
{\cal Z}_{3,k}
&=& {y_1^{4+k} \over (y_1 \bm y_2 )(y_1 \bm y_3 )} +{y_2^{4+k} \over (y_2 \bm y_1 )(y_2 \bm y_3 )} +
{y_3^{4+k} \over (y_3 \bm y_1 )(y_3 \bm y_2 )}\cr
 && \nonumber \\
&=& { y_1^{4+k} (y_2 - y_3 )+
y_2^{4+k} (y_3 - y_1 ) + y_3^{4+k} (y_1 - y_2 ) \over {\rm i}\,\sqrt{z(4+27z)}} \nonumber
\eea
and for general $g$
\be
{\cal Z}_{g,k} =\sum_{i=1}^g y_i^{2(g-1)+k} \prod_{j \neq i , j=1}^g {1\over y_i - y_j}
\nonumber\ee
i.e., (\ref{linealbis}).
Again the terms $y_i^{2(g-1)+k}$ for $i>1$ make nonperturbative in $1/k$ contributions, but the extra factor
$y_1^{2(g-1)}/ \prod_{j=2}^g ({y_1 - y_j})$ in the first term also makes perturbative in $1/k$ contributions
to the thermodynamic grant partition function $y_1^k$. In this sense, the periodic spectrum is ``closer" to the
thermodynamic limit than the lineal one.

\subsection{Thermodynamics in a weak potential trap and in free space}

A $1$-dimensional harmonic trap potential acts as a ``box," and taking the strength of the potential to zero should recover the
free space result. However, the limit is non trivial and has to be taken appropriately. In particular, the
density of states in the potential will not converge to that of free space. E.g., the harmonic
oscillator density of states is constant and equal to $1/\omega$, while the free density of states in a large
flat box of length $L$ is $L /\pi\sqrt{2\epsilon}$ (we put the mass of the particle and $\hbar$ to 1).
Clearly the two have different energy dependence and will never converge in the limit $\omega \to 0$, $L\to \infty$.
Thermodynamics ``sees" the effect of the potential at large distances.

Assume the system is in a potential $V(x/\lambda)$, with $\lambda$ a scaling parameter. Taking $\lambda\to \infty$
speads the potential and corresponds to the infinite volume limit. In that limit, the potential becomes
increasingly smooth, as its derivatives scale like $1/\lambda$. For large enough $\lambda$ we can cut the
system into intervals of size $L$ and take $L$ large enough to contain a macroscopically large number of particles
but small enough for the potential inside it to be considered constant. The full grand potential $\ln {\cal Z}$ of the
system will be the sum of the grand potentials $\ln {\cal Z}_n$ of the part of the system in each interval
$nL<x<(n+1)L$,
\be\nonumber
\ln {\cal Z (\lambda,\mu)} = \sum_n \ln {\cal Z}_n (\lambda,\mu) ~,~~~ nL <x <(n+1) L
\ee
where we displayed the dependence on $\lambda$ and the chemical potential $\mu$.
Since the potential is almost constant in each interval, $\ln {\cal Z}_n$ is the grand potential in free space
of length $L$, and the constant potential simply shifts the chemical potential of the system. Calling
$\ln {\cal Z}_{\text{free}} (L,\mu)$ the free grand potential, we have 
\be\nonumber
\ln {\cal Z (\lambda,\mu)} \simeq \sum_n \ln {\cal Z}_{\text{free}}\big(L,\mu-V(nL/\lambda)\big)
\ee
and in the linit $\lambda \to \infty$, turning the sum to an integral,
\be
{1\over \lambda}\ln {\cal Z (\lambda,\mu)} = {1\over L}\int dx \ln {\cal Z}_{\text{free}}\big(L,\mu-V(x)\big)
\label{befree}\ee
The above provides an integral relation between ${\cal Z}$ and ${\cal Z}_\text{free}$. Since $\ln\bb {\cal Z}_\text{free}$
is extensive it scales as $L$, and so the RHS of (\ref{befree}) is independent of $L$.
Likewise, $\ln {\cal Z (\lambda,\mu)}$ will scale as $\lambda$, and the LHS is independent of $\lambda$ in
the limit $\lambda \to \infty$. { This implies the relation between particle numbers
\be
{1\over \lambda} N(\lambda,\mu) ={1\over L} \int dx N_\text{free} \big(N,\mu-V(x)\big)
\label{nfree}\ee}
and expanding the grand potentials in the fugacity $z=e^{\beta \mu}$ we obtain the relation of the cluster coefficients
\be
{1 \over \lambda}b_n (\lambda) ={1\over L} b_{n,\rm{free}} (L) \int dx \, e^{-n\beta V(x)}
\label{al}\ee
This result can also be obtained from the relation (\ref{genb}) derived in Section (2). Indeed, in the thermodynamic limit,
the single-particle partition function in the RHS of (\ref{genb}) can be well approximated by its semiclassical
expression
\be\nonumber
\sum_i e^{-n\beta \epsilon_i} = \int {dx\, dp \over 2\pi} e^{-n\beta H(p,x)} =
\int {dx\, dp \over 2\pi} e^{-n\beta (p^2/2 +V(x))} ={1\over \sqrt{2\pi n\beta}} \int dx e^{-n\beta V(x)}
\ee
and applying it to a potential $V(x)$, and to a square well of width $L$, we reproduce (\ref{al}). In particular,
for either inclusion ($g<0$) or exclusion ($g>0$), we obtain for a free system
\be\nonumber
b_{n,\text{free}} = L\, {c_n (g) \over \sqrt{2\pi n \beta}} 
\ee\nonumber
and for a harmonic potential trap $V(x) = \half \omega^2 x^2$
\be\nonumber
b_n (\omega) = {c_n (g) \over \beta n \omega}
\ee
that is, the bosonic results times the statistics factor $c_n (g)$.

{ The fact that the $1$-dimensional  harmonic trap for inclusion statistics
manifests condensation, while the free system does not, can be derived from (\ref{nfree}): for $z \to z_\text{max}=(g-1)^{g-1}/g^g$, the
density of particles for the free system diverges at $x=0$ but remains finite for other $x$, and the integral over $x$ is
finite, implying a maximal number of particles and condensation.

\end{document}